\documentclass[twocolumn,10pt]{IEEEtran}
\usepackage{bm,cite,float,amsmath,amssymb,amsthm}
\usepackage{multirow}
\usepackage{algorithm}
\usepackage[noend]{algpseudocode}
\usepackage{indentfirst}
\usepackage{xcolor}
\usepackage{makecell}
\usepackage{color}
\usepackage{bbding}
\usepackage{lipsum}
\usepackage{mathtools}
\usepackage{subfigure}
\usepackage{cuted}
\usepackage[caption=false,font=footnotesize]{subfig}

\makeatletter
\def\BState{\State\hskip-\ALG@thistlm}
\makeatother

\usepackage{amssymb}
\usepackage{amsmath}
\usepackage{graphicx}
\usepackage{cite}

\usepackage{balance}
\usepackage[utf8]{inputenc}

\usepackage{color,soul}

\usepackage{adjustbox}

\usepackage{verbatim}

\IEEEoverridecommandlockouts

\usepackage{graphicx,epstopdf}
\usepackage{epsfig}	
\usepackage{amsfonts,balance}
\usepackage{bbm}
\floatname{algorithm}{Algorithm}

\usepackage{lipsum}

\newtheorem{theorem}{Theorem}

\newtheorem{corollary}{Corollary}

\makeatletter
\def\ScaleIfNeeded{%
	\ifdim\Gin@nat@width>\linewidth \linewidth \else \Gin@nat@width
	\fi } \makeatother

\begin{document}
	
	%
	\title{Goal-Oriented Semantic Communications \\for 6G Networks}
	\author{Hui Zhou, Yansha Deng, Xiaonan Liu, Nikolaos Pappas,
		and Arumugam Nallanathan
		
		\thanks{Hui Zhou, and Yansha Deng are with the Department of Engineering, King’s College London, London, U.K. (e-mail: \{hui.zhou, yansha.deng\}@kcl.ac.uk) (Corresponding author: Yansha Deng).}
		\thanks{Xiaonan Liu, and Arumugam Nallanathan are with the School of Electronic Engineering and Computer Science, Queen
			Mary University of London, London, U.K. (e-mail: \{x.l.liu,  a.nallanathan\}@qmul.ac.uk).}
		\thanks{Nikolaos Pappas is with the Department of Computer and Information Science, Linköping University, Sweden (email:nikolaos.pappas@liu.se).}
		
	}
\maketitle



\begin{abstract}
Upon the arrival of emerging devices, including Extended Reality (XR) and Unmanned Aerial Vehicles (UAVs), the traditional communication framework is approaching Shannon's physical capacity limit and fails to guarantee the massive amount of transmission within latency requirements. By jointly exploiting the context of data and its importance to the task, an emerging communication paradigm shift to semantic level and effectiveness level is envisioned to be a key revolution in Sixth Generation (6G) networks. However, an explicit and systematic communication framework incorporating both
semantic level and effectiveness level has not been proposed yet. In this article, we propose a generic goal-oriented semantic communication framework for various tasks with diverse data types, which incorporates both semantic level information and effectiveness-aware performance metrics. We first analyze the unique characteristics of all data types, and summarise the semantic information, along with corresponding extraction methods. We then propose a detailed goal-oriented semantic communication framework for different time-critical and non-critical tasks. In the goal-oriented semantic communication framework, we present the goal-oriented semantic information, extraction methods, recovery methods, and effectiveness-aware performance metrics. Last but not least, we present a goal-oriented semantic communication framework tailored for Unmanned Aerial Vehicle (UAV) control task to validate the effectiveness of the proposed goal-oriented semantic communication framework.

\end{abstract}



%
\maketitle

\section{Introduction}

Inspired by Shannon’s classic information theory, Weaver and Shannon proposed a more general definition of a communication system involving three different levels of problems, namely, (i) the bits conveying
information should be reliably transmitted to the recipient (the technical problem); (ii) the context conveyed by the transmitted bits should accurately reflect the intentions of the sender (the semantic problem); and (iii) the conduct or action
of the system in response to communications should be effective in accomplishing
a desired task (the effectiveness problem) \cite{Shannon1949}. The first level of communication, which is the transmission of bits, has been well studied and realized in conventional communication systems based on Shannon’s technical framework.
However, with the massive deployment of emerging devices, including Extended Reality (XR) and Unmanned Aerial Vehicles (UAVs), diverse tasks with stringent requirements pose critical challenges to traditional communications, which are already approaching the Shannon physical capacity limit. This imposes the Sixth Generation (6G) network towards a communication paradigm shift to semantic level and effectiveness level by exploiting the context of data and its importance to the task. It is noted that the significance and importance of information evaluates the importance of extracted semantic information in accomplishing a specific task and is closely coupled with the considered task.

Initial works on ``semantic communications'' have mainly focused on identifying the content of the traditional text and speech \cite{ Luo2022}, and the information freshness, i.e., age of information (AoI) \cite{pappas2023age} as a semantic metric that captures the timeliness of the information. However, these cannot capture the data importance sufficiently of achieving a specific task. In \cite{kountouris2021semantics}, a joint design of information generation, transmission, and reconstruction was proposed. Although the authors explored the benefits of including the effectiveness level in \cite{popovski2020semantic}, an explicit and systematic communication framework incorporating both semantic level and effectiveness level has not been proposed yet. There is an urgent need for a unified communication framework aiming at task-oriented performances for diverse data types.





%


%

Motivated by this, in this paper, we propose a generic goal-oriented semantic communication framework, which jointly considers the semantic level information about the data context and effectiveness-aware performance metrics about data importance for different tasks with various data types. The main contributions of this paper are: 
\begin{enumerate}
\item We first present the existing semantics for traditional text, speech, image, and video data types. More importantly, we analyze the unique characteristics of emerging data types, including $360^\circ$ video, sensor, haptic, and machine learning models, and propose corresponding semantics definition and extraction methods in Section~II.
\item We then propose a generic goal-oriented semantic communication framework for typical time-critical and non-critical tasks, where semantic level and effectiveness level are jointly considered. Specifically, by exploiting the unique characteristics of different tasks, we present goal-oriented semantic information, their extraction and recovery methods, and effectiveness-aware performance metrics to guarantee the task requirements in Section~III.
\item To demonstrate the effectiveness of our proposed goal-oriented semantic communication framework, we present the goal-oriented semantic communication solution tailored for Unmanned Aerial Vehicle (UAV) control and analyze the results in Section IV.
\end{enumerate}

\section{Semantic Information Extraction}

To exploit the context of the data for transmission, the challenge lies in designing the algorithm to identify and then extract the semantic information from each data type based on its unique characteristics. It is noted that each data type needs a customized semantic information extraction algorithm to exploit its characteristics fully. Therefore, in this section, we focus on analyzing the characteristics of both traditional and emerging data types in the 6G network and summarizing the semantic information definition with corresponding extraction methods, as shown in Table~\ref{feature_summarization}.


\begin{table}[!h]
\caption{Semantic information extraction of different data types}
\begin{center}
	\begin{tabular}{|c|c|c|}
		\hline
		\textbf{Data Type}  &Semantic Information &\makecell[c]{Semantic Information\\ Extraction Method}\\
		\hline			
		Text &Embedding&BERT \\
		\hline
		Speech &Embedding&BERT \\
		\hline
		Image &Edge, Corner, Blob, Ridge&SIFT, CNN \\
		\hline
		Video &Temporal Correlation&\makecell[c]{CNN} \\
		\hline
		\makecell[c]{$360^\circ$ Video }&FoV&\makecell[c]{Biological Information\\ Compression}\\
		\hline
		Haptic Data &JND&Web's Law\\
		\hline
		\makecell[c]{Sensor and\\ Control Data}&Freshness& AoI\\
		\hline
	\end{tabular}
	\label{feature_summarization}
\end{center}
\end{table}

\subsection{{{Speech and Text}}}
For one-dimensional speech signal, the speech-to-text conversion can be first performed by speech recognition. With the extracted text information, the famous embedding extraction method, i.e., Bidirectional Encoder Representations from Transformers (BERT) can be applied to extract embedding as typical semantic information, which represents the words, phrases, or text as a low-dimensional vector \cite{Luo2022}. However, during the speech-to-text conversion process, the timbre and emotion conveyed in the speech may be lost.




\subsection{{Image and Video}}
As a two-dimensional data type, the image geometric structures, including edges, corners, blobs, and ridges, can be identified as typical semantic information, where the Convolutional Neural Network (CNN) has shown stronger capability to extract complex geometric structures with its matrix kernel \cite{Zhao2019, Minaee2022}.  Since video combines two-dimensional images with an extra time dimension, the temporal correlation between adjacent frames can be identified as important semantic information, where the static background can be ignored during transmission.






\subsection{{$360^\circ$ Video}}
The $360^\circ$ rendered video is a new data type in emerging XR applications. The most important semantic information is identified as human field-of-view (FoV), which occupies around one-third of the $360^\circ$ video and only has the highest resolution requirement at the center \cite{Liu_vr}. In this case, biological information, such as retinal foveation and ballistic saccadic eye movements can be leveraged for semantic information extraction. Therefore, biological information compression methods have been utilized to extract the semantic information, where retinal foveation and ballistic saccadic eye movements are jointly considered to optimize the semantic information extraction process.


\subsection{{Haptic Data}}
Haptic data consists of two submodalities, which are tactile information and kinesthetic \cite{Antonakoglou2018,Li2022}. For tactile information, 
five major dimensions can be identified, which are friction, hardness perception, warmth conductivity, macroscopic roughness, and microscopic roughness.  
Kinesthetic information refers to the position/orientation of human body parts and external forces/torques applied to them. 
To reduce the redundant raw haptic data, Just Noticeable
Difference (JND) is identified as valuable semantic information to filter the haptic signal that cannot be perceived by the human, where Weber's law serves as an important semantic information extraction criterion. 


\subsection{Sensor and Control data}

Sensors are usually deployed to monitor the physical characteristics of the environment (e.g., temperature, humidity, or traffic) in a geographical area. The acquisition of data is transformed to status updates that are transmitted through a network to the destination nodes. Then, these data are processed to extract useful information, such as control commands or remote source reconstruction, that can be further utilized to predict the evolution of the initial source. The accuracy of the reconstructed data, either in control commands or in predicting the evolution, is directly related to the relevance or the semantic value of the data measurements. Thus, an important aspect is the generation of traffic and how it can be affected in order to filter only the most important samples so the redundant or less useful data will be eliminated to reduce potential congestion inside the network.

The AoI also has a critical role in dynamic control systems since it was shown that non-linear AoI and Value of Information (VoI) are paradigm shifts, and they can improve the performance of such systems. Furthermore, we have seen in early studies that the semantics of information (beyond timeliness) can provide further gains by reducing the amount of information that is generated and transmitted without degrading the performance.


\subsection{{Machine Learning Model}}
With the massive deployment of machine learning algorithms, machine learning-related model has been regarded as another important data type. 

\begin{itemize}
\item \textbf{Federated Learning (FL) Model:} The FL framework has been considered as a promising approach to preserve data privacy, where each participating device uploads the model gradients or model weights to the server and receives the global model from the server. 
\item \textbf{Split Learning (SL) Model:} Due to the limited computation capability of devices and heavy computation burden, SL has been proposed to split the neural network model between the server and devices, where the device executes the model up to the cut layer and sends the smashed data to the central server to execute the remaining layers. Then the gradient of the smashed data is transmitted back from the server to update the local model. 
\end{itemize}

However, it is noted that explicit semantic information definition for machine learning modes has not been proposed yet.

\section{Goal-Oriented Semantic Communication Framework}

\label{different_task}
\begin{figure}[!h]
\centerline{\includegraphics[scale=0.4]{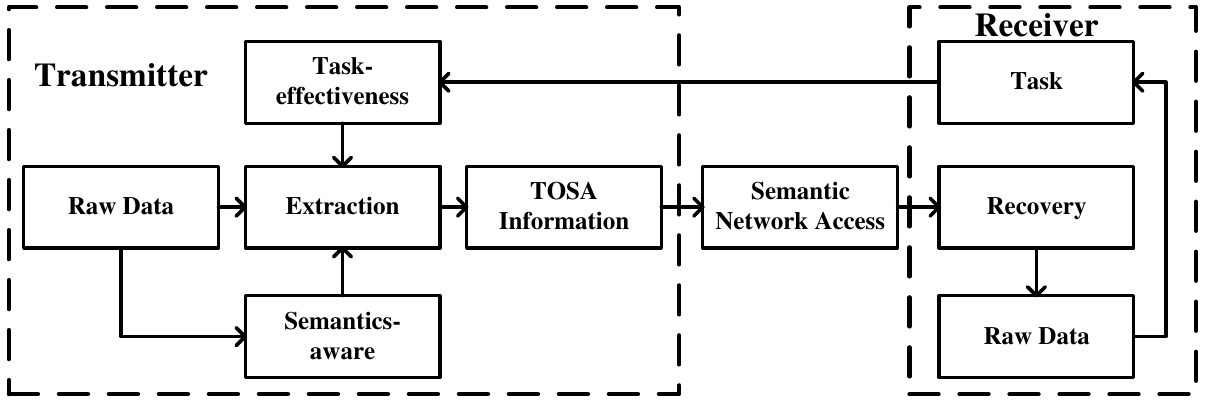}}
\caption{Proposed goal-oriented semantic communication framework.}
\label{Proposed_framework}
\end{figure}

In this section, we propose a generic goal-oriented semantic communication framework incorporating both semantic level and effectiveness level as shown in Fig.~\ref{Proposed_framework}, which consists of goal-oriented semantic information, its extraction and recovery methods, and effectiveness-aware performance metrics. It is noted that the goal-oriented semantic information is extracted based on the semantics of the raw data and feedback of the task execution effectiveness, which exploits the context of data and its importance to the task. In the following, we will illustrate the implementation of task-effectiveness feedback component for each specific task as summarized in Table~\ref{Tasks_summarization}.

\begin{figure*}[!h]
\centerline{\includegraphics[scale=0.7]{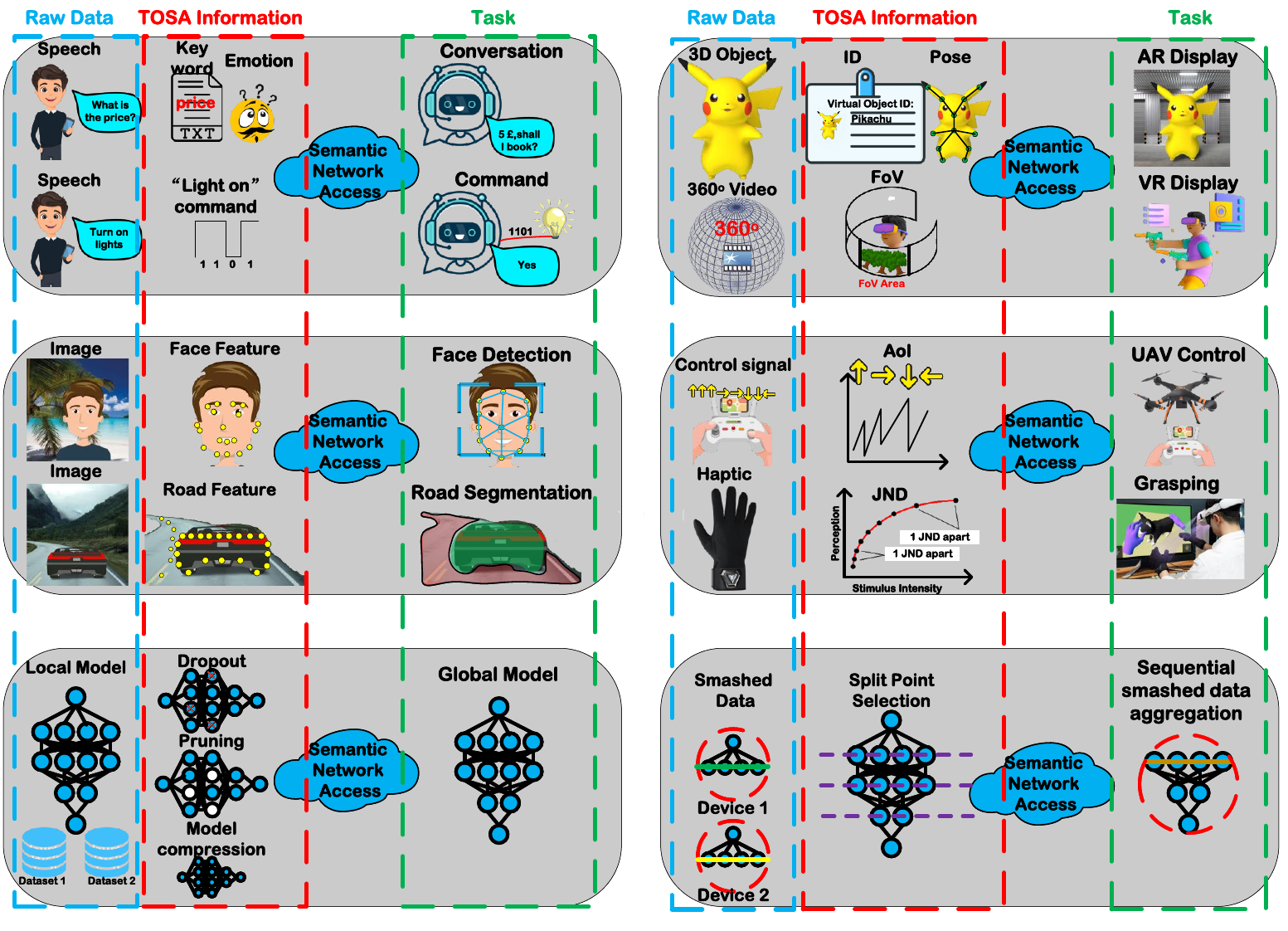}}
\caption{Goal-oriented semantic communication framework for different tasks with diverse data types.}
\label{TOSA_framework_fig}
\end{figure*}

\begin{table*}[!h]
\caption{Goal-oriented semantic Communication Framework Summarization of different tasks}
\begin{center}
	\begin{tabular}{|c|c|c|c|c|c|}
		\hline
		\textbf{Data Type} & Task & Communication Entity& Recovery& Latency Type & \makecell[c]{Effectiveness Level\\ Performance Metrics}\\
		\hline			
		\multirow{1}{*}{\makecell[l]{Speech}}&\makecell[c]{Speech Recognition} &  Human-Machine &Yes/No&Non-critical &\makecell[l]{$\bullet$F-measure\\$\bullet$Accuracy\\$\bullet$BLEU\\$\bullet$Perplexity}  \\
		\hline
		\multirow{5}{*}{\makecell[l]{Image}}& \makecell[c]{Object Detection} & Machine-Machine &No&Non-critical &  \makecell[l]{$\bullet$IoU\\$\bullet$mAP\\$\bullet$F-measure\\$\bullet$MAE} \\
		\cline{2-5}
		& \makecell[c]{Semantic Segmentation} & Machine-Machine&Yes & Critical &\makecell[l]{$\bullet$IoU\\$\bullet$Pixel Accuracy\\$\bullet$MPA\\$\bullet$Latency}\\
		\hline
		\multirow{5}{*}{\makecell[c]{$360^\circ$ Video}} & \makecell[c]{Display in AR}& Machine-Human& Yes& Critical &\makecell[l]{$\bullet$PSNR\\$\bullet$SSIM\\$\bullet$Alignment Accuracy}  \\
		\cline{2-5}
		& \makecell[c]{Display in VR} & Machine-Human& No& Critical & \makecell[l]{$\bullet$PSNR\\$\bullet$SSIM\\$\bullet$Timing Accuracy\\$\bullet$Position Accuracy}\\
		\hline
		\makecell[c]{Haptic Data} & Grasping and Manipulation & Machine-Human&No & Critical& \makecell[l]{$\bullet$SNR\\$\bullet$SSIM}\\
		\hline
		Sensor& Networked control systems & Machine-Machine& No& Critical& \makecell[l]{$\bullet$LGQ}\\
		\hline
		\multirow{5}{*}{\makecell[c]{ML\\Model}}& Federated Learning & Machine-Machine &\rule[0pt]{1cm}{0.1em}& Critical/Non-Critical& \makecell[l]{$\bullet$Latency\\$\bullet$Reliability\\ $\bullet$Convergence Speed\\$\bullet$Accuracy} \\
		\cline{2-6}
		& Split Learning & Machine-Machine &\rule[0pt]{1cm}{0.1em}& Critical/Non-Critical & \makecell[l]{$\bullet$Latency\\ $\bullet$Reliability \\$\bullet$Convergence Speed \\$\bullet$Accuracy}\\
		\hline
	\end{tabular}
	\label{Tasks_summarization}
\end{center}
\end{table*}

\subsection{One-hop Task}
We consider one-hop tasks with a single link transmission in this section for typical time-critical and non-critical tasks as shown in Fig.~\ref{TOSA_framework_fig}, where each communication entity can be either human or machine as summarized in Table~\ref{Tasks_summarization}.

\subsubsection{{Speech Recognition}} The speech recognition task can be further divided into conversation-type tasks (e.g., human inquiry) and command-type tasks (e.g., smart home control) depending on the speech content. The conversation-type task focuses on understanding the intent, language, and sentiment to provide humans with free-flow conversations. The command-type task focuses on parsing the specific command over the transmitted speech and then controlling the target device/robot. 

In the conversation-type task, the goal-oriented semantic information can be keywords and emotions. The device can obtain the goal-oriented semantic information by transforming the speech signal into text and extracting keywords and emotions via BERT. Then, the server recovers the text via the transformer decoder. In the command-type task, the goal-oriented semantic information can be the binary command, and the device can directly parse the speech signal and obtain the binary command signal for transmission, where no recovery is needed on the receiver side. The effectiveness-aware performance metrics include F-measure, accuracy, bilingual evaluation understudy (BLEU), and perplexity, where user satisfaction should also be considered.  



\subsubsection{{Object Detection and Semantic Segmentation}}  
Object detection and semantic segmentation are two emerging image processing tasks \cite{Zhao2019, Minaee2022},  where the captured images are required to be transmitted to the central server for processing. However, the semantic segmentation task, e.g., road segmentation in autonomous driving applications, imposes stringent latency and reliability requirements due to road safety issues. This is because the vehicles must react instantly to the rapidly changing environment.

For the time non-critical object detection task, e.g., face detection, goal-oriented semantic information can be the face feature extracted via CNN. After being transmitted to the central server, the regions with CNN features (R-CNN) can be applied to perform face detection. For the time-critical road segmentation task, one possible solution is to identify the region of interest (ROI) features, i.e., road, as the goal-oriented semantic information, and crop the images via region proposal algorithms. Then, the central server can perform image segmentation via mask R-CNN. Both tasks can be evaluated via effectiveness-aware performance metrics, including Intersection over Union (IoU), mean average precision (mAP), F-measure, and mean absolute error (MAE). It is noted that road segmentation in the autonomous driving application can be evaluated via pixel accuracy, and mean pixel accuracy (MPA). However, the trade-off between accuracy and latency remains an important challenge to solve.


\subsubsection{{Display in Extended Reality}}



In the Augmented Reality (AR) display task, the central server transmits the rendered 3D model of a specific virtual object to the user. It is noted that the virtual object identification and its pose information related to the real world is the key to achieving alignment between virtual and physical objects. Therefore, the virtual object identification and pose information can be extracted as goal-oriented semantic information to reduce the data size. Then, by sharing the same 3D model library, the receiver can locally reconstruct the 3D virtual object model based on the received goal-oriented semantic information. To evaluate the 3D model transmission, Peak Signal-to-Noise Ratio (PSNR) and Structural Similarity (SSIM) can be adopted as effectiveness-aware performance metrics. However, how to quantify the alignment accuracy among the virtual objects and physical objects as a performance metric remains to be solved.




In the VR display task, the central server transmits virtual $360^\circ$ video streaming to the user. To avoid the transmission of the whole $360^\circ$ video, the central server can predict the eye movements of the user and extract the corresponding FoV as goal-oriented semantic information. Apart from the PSNR and SSIM mentioned in AR, timing accuracy and position accuracy are also important effectiveness-aware performance metrics to avoid cybersickness including: 1) initial delay: time difference between the start of head motion and that of the corresponding feedback; 2) settling delay: time difference between the stop of head motion and that of the corresponding feedback; 3) precision: angular positioning consistency between physical movement and visual feedback in terms of degrees; and 4) sensitivity: capability of inertial sensors to perceive subtle motions and subsequently provide feedback to users.


\subsubsection{{Grasping and Manipulation}}
Haptic communication has been incorporated by industries to perform grasping and manipulation, where the robot transmits the haptic data to the manipulator. The shape and weight of the objects to be held are measured using cutaneous feedback derived from the fingertip contact pressure and kinesthetic feedback of finger positions, which should be transmitted within stringent latency requirements to guarantee industrial operation safety.

Due to the difficulty in supporting massive haptic data with stringent latency requirements, JND can be identified as important goal-oriented semantic information to ignore the haptic signal that cannot be perceived by the manipulator. Two effectiveness-aware performance metrics including SNR and SSIM have been verified to be applicable to vibrotactile quality assessment.

\subsubsection{Control}
In networked control systems (NCS), typically, multiple sensors measure the system state of their control processes and transmit the generated data over a resource-limited shared wireless network. These data usually are enqueued and then transmitted over unreliable channels that cause excessive delays resulting in outdated or even obsolete for decision-making. Therefore, data freshness and importance are extracted as goal-oriented semantic information via AoI and VoI to guarantee the timing requirement, respectively \cite{TimingProcIEEE}. A typical effectiveness-aware performance metric that is used to minimize is the Linear Quadratic Gaussian (LQG) cost function, and usually the lower the value of the LQG function the higher the quality of control (QoC).

\subsubsection{{Machine Learning}}
In the following, we propose the goal-oriented semantic communication for two distributed ML models, i.e., FL and SL, where the performance metrics of FL and SL tasks are the convergence of these learning algorithms. Specifically, the goal-oriented semantic communication for FL and SL tasks is designed to transmit the most important weights/ neuron parameters of FL and SL without sacrificing the convergence performance.

\paragraph{Federated Learning} For the time non-critical tasks, the goal of FL is to guarantee a high learning accuracy without latency constraints. Traditional loss functions, such as mean square error (MSE), MAE, and cross-entropy, can be directly used as effectiveness-aware performance metrics. However, for time-critical tasks, the goal of FL is to balance the trade-off between learning accuracy, communication latency, and computation latency. The effectiveness-aware performance metrics are loss functions with latency constraints. Time-critical tasks bring 
communication challenges, and communication-efficient FL should be designed to decrease the model size to satisfy latency constraints via federated dropout, federated pruning, and model compression. 

Federated dropout randomly drops neurons during the training phase, which decreases communication and computation latencies and slightly improves learning accuracy. However, during the testing phase, the whole learning model is transmitted between the server and devices. Federated pruning permanently removes neurons in either or both training and testing phases. The pruning ratio should be carefully designed to guarantee learning accuracy, and extra computation latency is required to calculate the importance of parameters. Thus, how to design federated pruning methods with low computation complexity needs to be investigated. Model-compression schemes decrease the model size via sparsification or
quantization. However, these methods slightly decrease the convergence rate and achieve a modest accuracy (about 85$\%$). Thus, how to design a model compression algorithm with high learning accuracy still needs to be investigated.

\paragraph{Split Learning} In SL, the smashed data and its gradient associated with the cut layer are the extracted task-oriented information and transmitted between the server and devices, where no model recovery is required. When multiple devices exist in SL, all devices interact with the edge server sequentially, resulting in high training latency. For time non-critical tasks, the goal of SL is to achieve high learning accuracy without latency constraints and the effectiveness-aware performance metrics are the same as that of the FL. However, for time-critical tasks, the SL cannot guarantee the requirement of low latency because of its sequential training pattern.

To adapt the SL to time-critical tasks, splitfed learning (SFL) \cite{thapa2022splitfed} and hybrid split and federated learning (HSFL) \cite{chengzhipeng2022} are proposed. The effectiveness-aware performance metrics of the SFL and HSFL are learning accuracy and training latency. However, SFL and HSFL assume that the model is split at the same cut layer and the server-side model is trained in a synchronous mode. Splitting at the same cut layer leads to asynchronization of device-side model training and smashed data transmission. Thus, how to select an optimal split point and deal with the asynchronization of SL remain important challenges to solve. Also, different split points can result in different smashed data. Thus, how to merge these smashed data in the server-side model should be considered.



\subsection{Chain Task}
In this section, we analyze more complicated but practical chain tasks including XR-aided teleoperation and chain of control, where multiple entities cooperate through communication links to execute the task. 

\subsubsection{XR-aided Teleoperation}

To implement a closed-loop XR-aided teleoperation system, the wireless network is required to support mixed types of data traffic, which includes control and command (C\&C) transmission, haptic information feedback transmission, and rendered $360^{\circ}$ video feedback transmission \cite{xr-aided}. As XR-aided teleoperation task relies on both parallel and consecutive communication links, how to guarantee the cooperation among these communication links to execute the task is of vital importance. Specifically, the parallel visual and haptic feedback transmissions should be aligned with each other when arriving at the manipulator, and consecutive C\&C and feedback transmissions should be within the motion-to-photon delay constraint, which is defined as the delay between the movement of the user's head and the change of the VR device's display reflecting the user's movement. Either violation of alignment in parallel links or latency constraint in consecutive links will lead to a break in presence (BIP) and cybersickness. Therefore, both parallel alignment and consecutive latency should be quantified into effectiveness-aware performance metrics to guarantee the success of XR-aided teleoperation. Moreover, due to the motion-to-photon delay, the control error between the expected trajectory and the actual trajectory will accumulate along with the time, which may lead to task failure. Hence, how to alleviate the accumulated error remains an important challenge that needs to be solved.

\subsubsection{Chain of control}
In the scenario of a swarm of (autonomous) robots where they need to perform a collaborative task (or a set of tasks) within a deadline over a wireless network, an effective communication protocol that takes into account the peculiarities of such a scenario is needed. Considering the simple case of two robots, let's say Robot A and Robot B that are communicating through a wireless network and they are not collocated. Robot A controls remotely Robot B such that to execute a task and the outcome of that operation will be fed to Robot A for performing a second operation to send the outcome back to Robot B. All this must happen within a strict deadline. The amount of information that is generated, transmitted, processed, and sent back can be very large with the traditional information agnostic approach. On the other hand, if we take into account the semantics of information and the purpose of communication, we change the whole information chain, from its generation point until its utilization. Therefore, defining goal-oriented semantic metrics for the control loop and communication between a swarm of (autonomous) robots is crucial and it will significantly reduce the amount of information leading to a more efficient operation.

\section{Case Study}

To validate the effectiveness of our proposed goal-oriented semantic communication framework,  we present the goal-oriented semantic communication solution tailored for UAV control task in this section. Specifically, the ground control station(GCS) periodically sends the C\&C signal to UAV user equipment(UAV-UE) via the terrestrial base station, where the UAV-UE flies in a circular horizontal disk. It is noted that UAV C\&C signals are usually sent periodically with consecutively redundant information, i.e., the same command will be sent multiple times continuously, which may easily lead to traffic congestion in traditional communication.

To solve the problem above, VoI and AoI are utilized to quantify the similarity between consecutive
C\&C signals and freshness of the C\&C signal, respectively \cite{Xu2023}. This is because the newly generated C\&C signal with new C\&C content is more important to the UAV-UE control task. Without transmitting every C\&C signal, the deep reinforcement learning (DRL) algorithm is utilized to maximize the goal-oriented semantic information (i.e., AoI and VoI), where only the important C\&C signals will be transmitted to the UAV-UE. Taking one example, the generated YAW C\&C signals can be $[30^{\circ}/s, 30^{\circ}/s, 30^{\circ}/s, 0^{\circ}/s, 0^{\circ}/s, 0^{\circ}/s]$. In this setting, all C\&C signals will be transmitted to UAV-UE in traditional communication. However, only $[30^{\circ}/s, 0^{\circ}/s]$ will be transmitted to the UAV-UE in goal-oriented semantic communications, which reduces over $60\%$ of data transmission at the base station.

\begin{figure}[!h]
\centering
\subfigure[Transmission times versus different consecutive repeat times $k$]
{\includegraphics[scale=0.35]{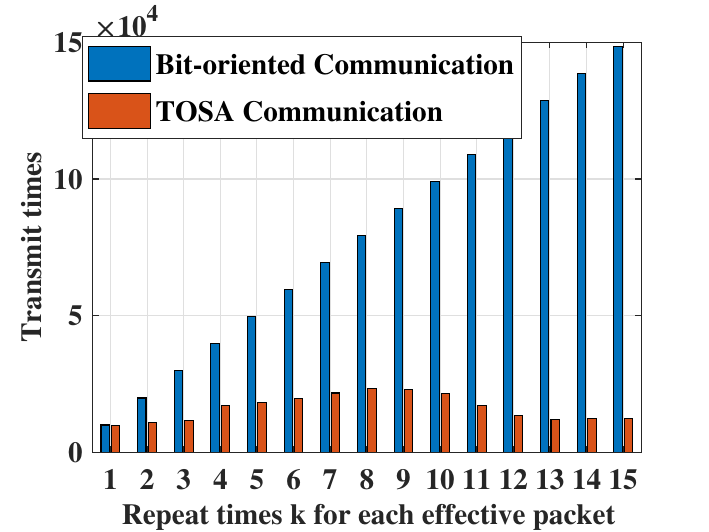}} 
\centering
\subfigure[Effective transmission rate versus various repeat times $k$]
{\includegraphics[scale=0.35]{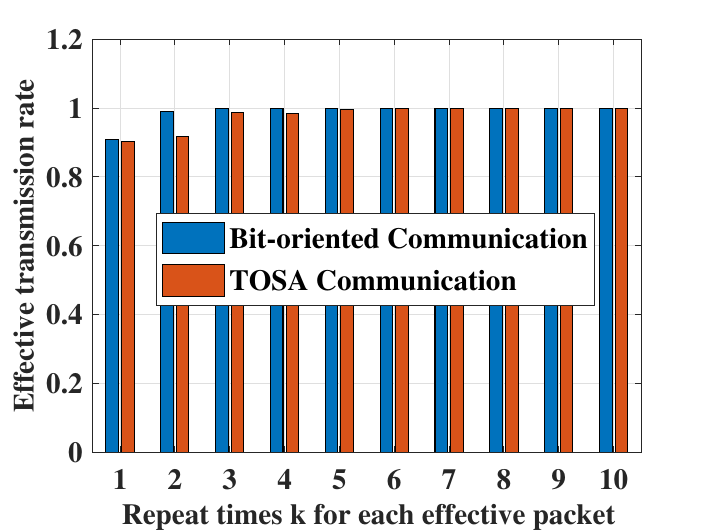}}
\caption{Transmission times and effective transmission rate.}
\label{TOSA_framework_fig_UAV}
\end{figure}

Fig.~\ref{TOSA_framework_fig_UAV} (a) presents the number of transmission times of conventional
	 communication and proposed goal-oriented semantic communication.
	With the increase of C\&C consecutively repetitive
	times $k$ (i.e., redundant information), the actual number of transmission times remains
	stable in our goal-oriented semantic communication framework, which is slightly larger than
	the number of effective packets due to transmission failure. It is noted that consecutive and repetitive C\&C packets are regarded as one effective C\&C packet.
	Therefore, compared to the conventional  communication, the
	BS learns to only transmit the important C\&C signal and drop
	the other less important C\&C signal. Fig.~\ref{TOSA_framework_fig_UAV} (b) illustrates the effective transmission rate, which is
	defined as the effective transmission times divided by the total
	effective packets. We can observe that our proposed
	goal-oriented semantic communication achieves almost the same effective
	transmission rate as traditional communication. It is noted that
	the negligible performance gain of traditional 
	comes from incredible resource consumption to transmit all
	the generated C\&C signals, which verifies that our proposed
	goal-oriented semantic communication can guarantee the effectiveness of the
	C\&C task execution with much lower resource consumption.



	\section{Conclusion}
	In this article, we propose a generic goal-oriented semantic communication framework incorporating both semantic and effectiveness levels for various tasks with diverse data types. We first identify the unique characteristics of all existing and new data types in 6G networks and summarize the semantic information with its extraction methods. To achieve task-oriented communications for various data types, we then present the corresponding goal-oriented semantic information, their goal-oriented semantic information extraction and recovery methods, and effectiveness-aware performance metrics for both time-critical and non-critical tasks. Importantly, our results demonstrate that our proposed goal-oriented semantic communication framework can be tailored for UAV C\&C signal transmission with much lower resource consumption. The paradigm shift towards the goal-oriented semantic communication design will flourish new research on task-driven, context and importance-aware data transmission in 6G networks, where the proposed unified goal-oriented semantic communication framework lays a solid foundation for diverse data types and tasks.
	

	%



	

	\ifCLASSOPTIONcaptionsoff
	\newpage
	\fi


	
	\bibliographystyle{IEEEtran}
	\bibliography{IEEEabrv,magazineV3}
	\vskip -2\baselineskip plus -1fil
	

	\vskip -2\baselineskip plus -1fil
	
	%
	
	%
	%
		
		
	
\end{document}